\documentclass[twocolumn,usenatbib]{mn2e}

\usepackage{graphicx}
\usepackage{subfigure}
\usepackage{amsmath}
\usepackage{txfonts}
\usepackage{color}
\usepackage[T1]{fontenc}

\usepackage{xcolor}
\usepackage[colorlinks]{hyperref}
\hypersetup{ colorlinks, linkcolor=blue, citecolor=blue, anchorcolor=blue }
\usepackage{natbib}
\bibpunct{(}{)}{;}{a}{}{,}

\title[Early UV signatures of SN Ia ejecta-companion interaction]{Early ultraviolet signatures from the interaction of Type Ia supernova ejecta with a stellar companion}
\author[Z. W. Liu]{Zheng-Wei Liu\thanks{E-mail:zwliu@ynao.ac.cn}, Takashi J. Moriya and Richard J. Stancliffe
                    \\
$^{1}$Argelander-Institut f\"ur Astronomie, Auf dem H\"ugel 71, D-53121, Bonn, Germany\\
}

\begin{document}

\date{Accepted 2015 September 3. Received 2015 August 24; in original form 2015 July 21}

%\pagerange{\pageref{firstpage}--\pageref{lastpage}} \pubyear{2002}

\maketitle

\label{firstpage}

\begin{abstract}

The progenitors of Type Ia supernovae (SNe Ia) are not yet fully understood. The two leading progenitor scenarios
are the single-degenerate (SD) scenario and the double-degenerate scenario. 
In the SD scenario, the collision of the SN Ia ejecta with its companion star is 
expected to produce detectable ultraviolet (UV) emission in the first few days after the 
SN explosion within certain viewing angles. A strong UV flash has recently been 
detected in an SN~2002es-like peculiar SN~Ia iPTF14atg by Cao et al., which 
is interpreted as evidence of an early-time UV signature due to SN ejecta interacting with its companion star,
supporting the SD scenario. In this paper, we present the expected luminosity distributions of early-time UV emission arising from 
SN Ia ejecta-companion interaction by performing binary population synthesis calculations for different progenitor systems 
in the SD scenario. 
Our theoretical predictions will be helpful for future early-time observations of SNe Ia to constrain their possible progenitors. 
Assuming the observed strong UV pulse of iPTF14atg was indeed produced by the SN ejecta-companion interaction, our population
synthesis model suggests 
that the progenitor system of iPTF14atg is most likely a red-giant donor binary system, and it is unlikely to have been a
 main-sequence or helium-star donor system.

\end{abstract}


\section{INTRODUCTION}
 \label{sec:introduction}

Type Ia supernovae (SNe~Ia) play a fundamental role in astrophysics. The use of SNe Ia 
as cosmic distance indicators has led to the discovery of the accelerating expansion of the Universe 
\citep{Ries98, Perl99, Leib08}. However, the nature of SN~Ia progenitors and the physics 
of the explosion mechanism remain mysterious \citep[see e.g.][for a review]{Hill00, Maoz14}. 
It is generally accepted that SNe~Ia arise from thermonuclear explosions of white dwarfs (WD) 
in binary systems \citep{Hoyl60, Finz67, Nomo82}. 
Depending on the nature of the companion star, the most favored progenitor models of SNe Ia are classified
into two general categories, the single-degenerate (SD) scenario \citep[e.g.][]{Whel73} and 
the double-degenerate (DD) scenario \citep{Iben84, Webb84}.

In the DD scenario, two carbon-oxygen (CO) WDs spiral in and merge due to gravitational wave radiation, 
resulting in a single object with a mass near the Chandrasekhar limit, which may then 
explode as an SN~Ia \citep{Iben84, Webb84}. Although the DD channel has been
suggested to lead to an accretion-induced collapse rather than a SN
Ia \citep{Nomo85, Saio98}, several hydrodynamical studies of mergers of DD systems
have concluded that some DD pairs can explode as SNe Ia \citep{Rasi95, Frye10, Pakm10, Pakm11b, 
Dan11, Moll14, Rask14}. In particular, \citet{Pakm12b}
showed that the violent merger of two CO WDs with masses of 0.9
$M_{\rm{\sun}}$ and 1.1 $M_{\rm{\sun}}$ can lead to events that
reproduce the observational characteristics of normal SNe~Ia. 

In the SD scenario, a WD accretes material from its
non-degenerate companion star, where the companion star could be either a hydrogen (H)-rich star such as
a main-sequence (MS), a slightly evolved sub-giant (SG) star or a 
red giant (RG) star, or a helium (He) star. When the mass of the WD approaches the
Chandrasekhar-mass limit, it explodes as an SN~Ia \citep[e.g.][]{Whel73, Nomo82, Han04}.  
In the H-accretion scenario, only a fairly
narrow range in the accretion rate above $10^{\rm{-7}}\ M_{\rm{\sun}}\, \rm{yr^{-1}}$ is allowed in order to attain stable H
burning on the surface of the WD, avoiding a nova explosion. This makes the SD scenario
difficult to explain the observed SN~Ia rate \citep{Han04, Mann05, Wang09, Wang12, Maoz12}. However, 
hydrodynamical simulations have shown that thermonuclear explosions of Chandrasekhar-mass WDs can 
reproduce some observational characteristics of either subluminous Type Iax supernovae (SNe Iax) \citep{Jord12a, Krom13, Krom15, Fink14},
or normal SNe Ia \citep{Game05, Seit13}, or overluminous SNe Ia \citep{Plew04, Meak09, Jord12b}, which depend on 
the ignition conditions of a Chandrasekhar-mass WD.

On the observational side, more and more evidence seems to favor the DD scenario, e.g., 
the non-detection of pre-explosion companion stars at the location of SNe Ia \citep{Li11, Bloo12}, the 
absence of H features in the nebular spectra of SNe Ia \citep{Leon07, Lund13, Lund15, Shap13}, the lack of radio and X-ray emission around peak
brightness \citep{Li11, Bloo12, Brow12, Chom12,  Hore12, Marg12},
and the absence of a surviving companion star in SN Ia remnants \citep{Kerz09, Scha12}.
All this evidence supports the DD scenario although most of them can also probably be explained by the
so-called ``spin-up/spin-down'' SD model \citep{Di11, Just11}. In the  ``spin-up/spin-down'' scenario, 
the fast spin of the WD can increase its critical explosion mass to be higher than
the maximum mass achieved by the WD, leading to the WD must spin down before it can explode \citep{Di11, Just11}. As a result,
the donor star might shrink rapidly before the WD explosion, because it would exhaust its H-rich envelope 
during a long spin-down timescale of the rapidly rotating WD until the SN Ia explosion. In such cases, 
the companion star probably evolves to be a WD when the SN Ia explodes, which may explain the lack of H in 
late spectra of SNe Ia and the absence of a surviving companion in the SN remnants \citep{Di11, Just11, Hach12}.
On the contrary, some observations \citep{Pata07, Ster11, Dild12} have 
detected signatures of H-rich circumstellar material (CSM) that are expected to exist around SNe Ia as 
the result of mass transfer from the companion, as well as WD winds \citep{Nomo82, Hach99}. This supports the 
SD scenario. However, this detected abundant CSM is in conflict with the missing radio signal for other SN Ia events, e.g. SN 
2011fe \citep{Chom12, Hore12}. Also, some studies suggest that CSM can also form in the DD scenario \citep{Shen13, Soke13}.

After the SN Ia explosion in the SD scenario, the SN ejecta expands freely for a few minutes to hours and then interacts with its 
non-degenerate companion star. As a result, the outer layers of the companion star are partially stripped and ablated while 
the star is shocked by the SN impact. Finally, the star survives the SN explosion and may show
some peculiar features such as overluminosity or heavy element enrichment \citep{Whee75, Mari00, Pakm08, Liu12, Liu13a, Liu13c, Pan12, Maed14}. 
The interaction of SN Ia ejecta with its companion star not only affects the star but also the SN itself. 
\citet[][]{Kase10} showed that the strong  emission arising from the 
collision of SN Ia ejecta with its companion star in a SD progenitor system should be detectable 
in the first few hours to days after the SN explosion under favorable viewing 
angles. This strong emission is expected to alter the SN light
curves at early times \citep{Kase10, Brow12, Cao15, Mori15, Olli15}. Compared to the brightness of the SN itself at early times, such strong emissions should be brightest in the 
ultraviolet (UV). Therefore, early UV observations are proposed as a direct way to test progenitor models of 
SNe Ia. Following the analytical method of \citet{Kase10}, analysis of observed early light-curves of 
SNe Ia has been carried out by different groups to look for evidence of shock emission and thus constrain possible 
progenitors of SNe Ia  \citep{Hayd10, Brow12, Brow12a, Cao15, Olli15}. Most interestingly, a strong UV flash 
after the SN explosion has recently been 
detected from early-time observations of a peculiar subluminous SN Ia iPTF14atg by \citet{Cao15}. They further interpreted this early UV 
flash as strong evidence of excess luminosity produced by SN ejecta interaction with its companion star, and suggested that some SNe 
Ia arise from the SD scenario \citep{Cao15}.

\begin{figure}
  \begin{center}
    {\includegraphics[width=\columnwidth, angle=360]{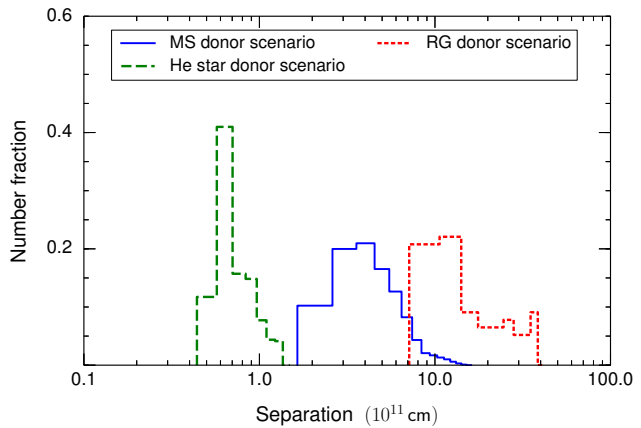}}
  \caption{Distributions of binary separations at the moment of SN explosion in different SD progenitor scenarios.}
\label{Fig:sep}
  \end{center}
\end{figure}

\begin{figure*}
  \begin{center}
    {\includegraphics[width=1.3\columnwidth, angle=270]{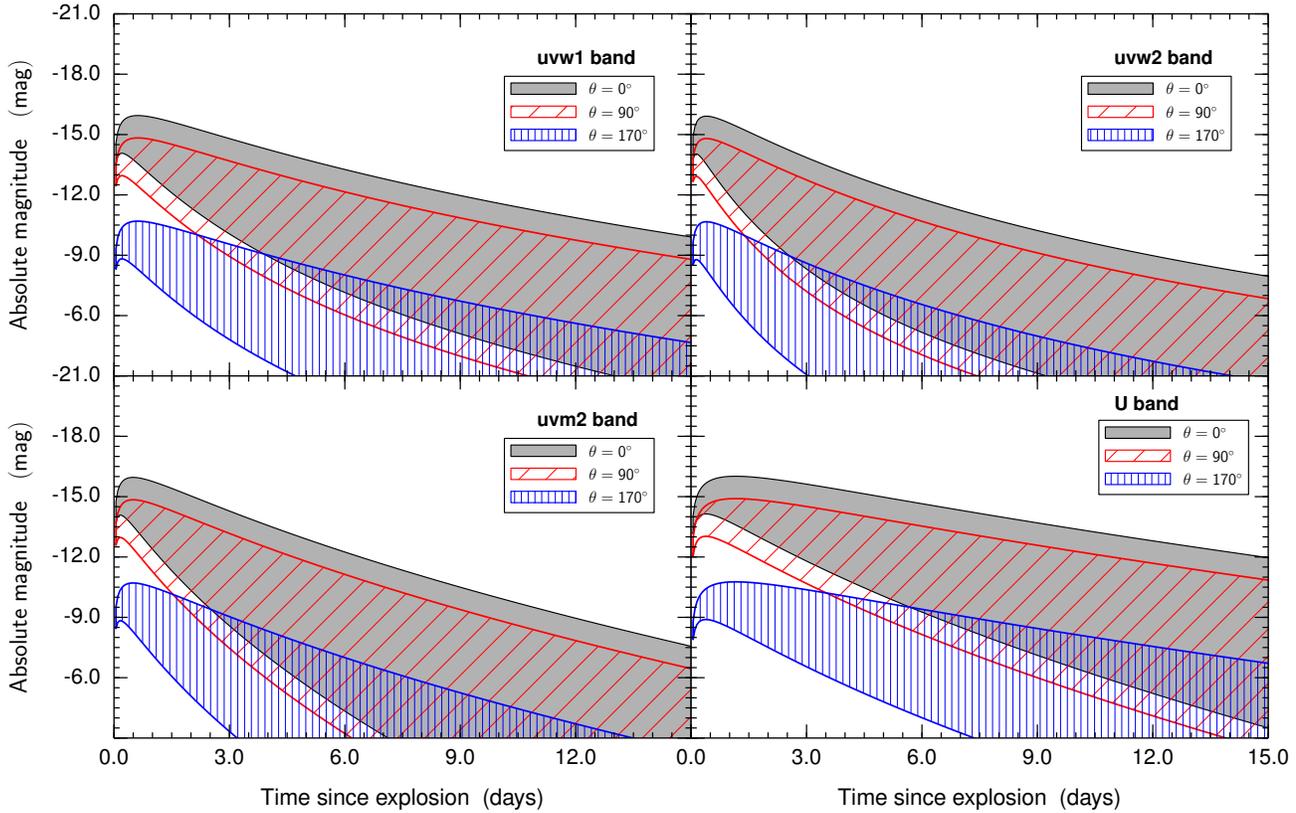}}
  \caption{Distributions of theoretical light curves predicted from the interaction between SN Ia ejecta and its MS companion star 
           in the four UV bands ($uvw1$, $uvm2$, $uvm2$ and $U$) of the $\textit{Swift}$ telescope. Here, $\theta$
           is the viewing angle, the maximum UV flux corresponds to a viewing angle of $\theta=0^{\circ}$ (looking down on the companion star).
           For a given viewing angle, each shaped region shows the covered magnitude range of our theoretical predictions with 
           the MS donor scenario at different epochs after the SN explosion. We assume the ejecta mass $M_{\rm{ej}}=1.4\,M_{\rm{\sun}}$ and
the SN explosion energy $E_{\rm{ej}}=1.0\times10^{51}\,\rm{erg}$.}
\label{Fig:1}
  \end{center}
\end{figure*}

In this work, we use binary population synthesis (BPS) models for different progenitor systems of SNe Ia in the SD 
scenario to obtain the progenitor properties at the moment of SN Ia explosion, e.g., the companion 
radius ($R$) and the binary separation ($a$). We then put these progenitor parameters into 
the analytical model of \citet{Kase10} to predict the 
distributions of early UV luminosity arising from the SN ejecta interacting with its companion star. Comparing our theoretical 
predictions with future early observations of SNe Ia will be helpful to constrain their possible progenitors. In addition, assuming 
the strong UV flash detected in iPTF14atg was indeed produced from the SN ejecta-companion interaction, we put further constraints 
on the progenitor system of iPTF14atg based on our BPS results.

The paper is organized as follows. In Section~\ref{sec:2}, we describe the BPS method used in this work. Theoretical early UV emission 
of SN ejecta-companion interaction are calculated in Section~\ref{sec:3}. In Section~\ref{sec:4}, 
we compare our predictions with the early-time observations of iPTF14atg and some discussions are presented. Our conclusions are 
summarized in Section~\ref{sec:5}.

\section{Binary population synthesis calculations}
\label{sec:2}

Adopting the method described in \citet{Han04}, we have performed BPS
calculations for the SD Chandrasekhar-mass scenario of SNe Iax by considering different types of companion
stars in \citet{Liu15}. With that BPS calculation, we have obtained the distributions of 
progenitor properties of different SD binary systems at the moment of SN explosion although we 
assumed all Chandrasekhar-mass CO WDs would lead to weak deflagration explosions 
and hence to SNe Iax  \citep{Krom13, Fink14} in that work. In the SD scenario, whether a Chandrasekhar-mass
CO WD would finally lead to a subluminous (i.e., Iax events), normal, or overluminous SN explosion, depends 
on the exact ignition conditions of the CO WD. However, as the WDs in our previous one-dimensional binary evolution 
calculations \citep{Liu15} were treated as point masses, it is impossible to determine the exact 
ignition conditions of the Chandrasekhar-mass CO WD. Therefore, we cannot distinguish whether the Chandrasekhar-mass CO WD undergoes
a delayed-detonation explosion that matches normal SNe Ia or a weak deflagration explosion of 
SNe Iax, and the progenitor properties at the moment of SN explosions are the same in our BPS calculations 
for the SD Chandrasekhar-mass scenario. Therefore, we directly use the data obtained from BPS 
calculations of \citet{Liu15} to calculate early UV emission of shocked ejecta of SNe Ia in the 
SD scenario in this work. Here, we only briefly describe 
our BPS method and basic assumptions, and we refer to \citet{Han04} for a detailed description 
about the BPS method used in this work.

\begin{figure*}
  \begin{center}
    {\includegraphics[width=1.3\columnwidth, angle=270]{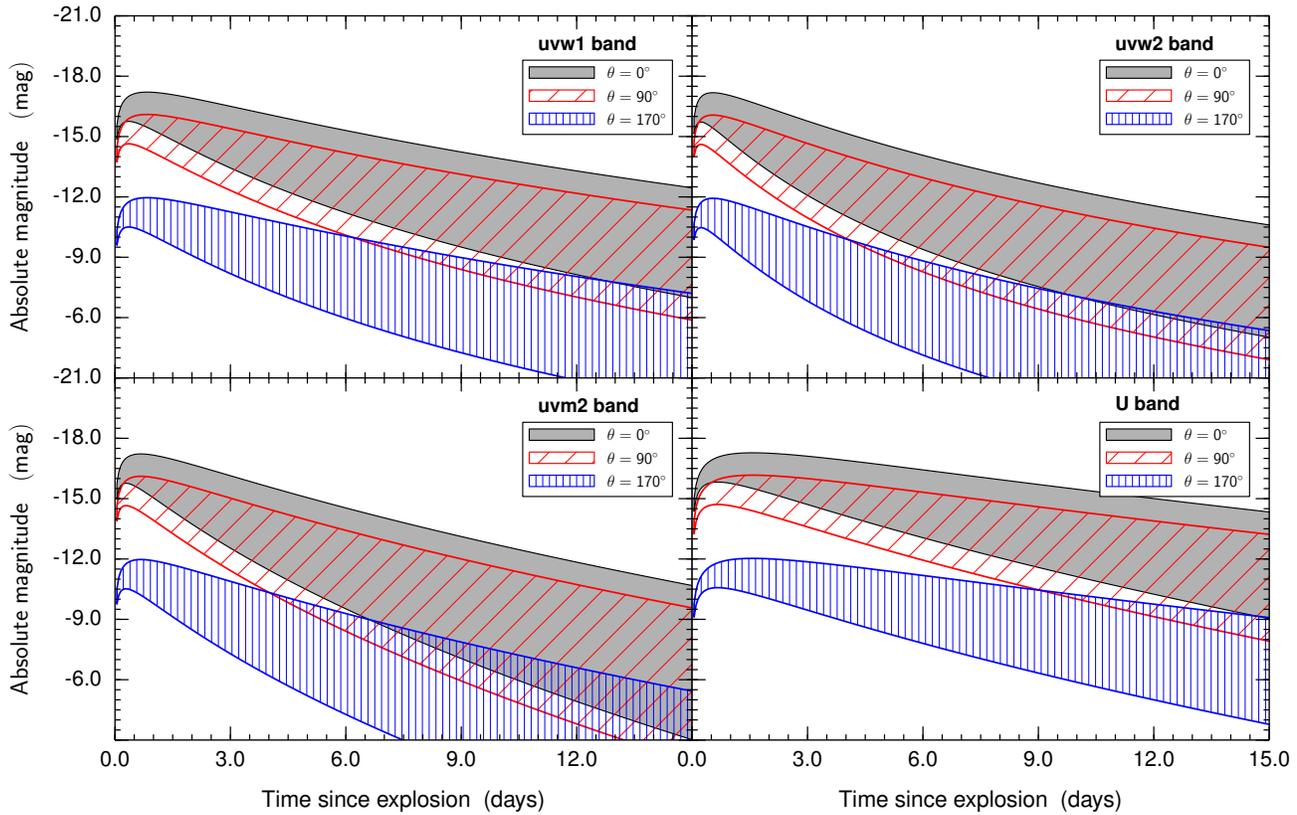}}
  \caption{As Fig.~\ref{Fig:1}, but for the RG donor Chandrasekhar-mass scenario.}
\label{Fig:2}
  \end{center}
\end{figure*}

We use the Cambridge stellar evolution code {\sc STARS} \citep{Eggl71, Eggl72, Eggl73} to trace the detailed 
binary evolution of a set of systems consisting of a CO WD and a MS or SG (which is called the MS donor 
scenario in the following), a RG, or a He star companion. 
Once the companion star fills its Roche lobe, the mass transfer occurs though Roche-lobe 
overflow (RLOF) which is treated by the prescription of \citet{Han00}. Once the WD has accreted the transferred H-rich 
or He-rich material from the companion star and increased its mass to near the Chandrasekhar-mass
limit ($\sim1.4\,M_{\sun}$), we then assume an SN Ia explosion occurs. The optically thick wind assumption of \citet{Hach96} is used to describe
the mass growth of a CO WD by the accretion of H-rich material from the donor star. The prescription of \citet{Kato04} 
is implemented for the mass accumulation efficiency onto the CO WDs when He-shell flashes occur. With a series 
of binary evolution calculations for various close WD binary systems, we determine the initial parameter space 
leading to SNe Ia in the orbital period--secondary mass (i.e. $\rm{log_{10}}$\,$P^{i}$--$M^{i}_{2}$ ) plane for various initial 
CO WD masses ($M^{i}_{\rm{WD}}$) in the MS, RG, and He star donor channel.

Furthermore, to obtain the distributions of progenitor properties of binary systems at the moment of SN explosion 
in the SD scenario, we use a rapid population synthesis code \citep{Hurl00, Hurl02} to 
perform a detailed Monte Carlo simulation. When a binary system evolves to a CO WD + MS (SG, RG, or He star) 
system and achieves the beginning of the RLOF phase, whether this binary system can lead to 
a SN Ia explosion is determined by checking if it is located in the 
SN Ia production regions in the plane of ($\rm{log_{10}}$\,$P^{i}$, $M_{2}^{i}$) for its $M_{\rm{WD}}^{i}$ based on our 
detailed binary evolution. Then, progenitor properties at the moment of SN explosion are obtained by interpolation in the three-dimensional 
grid ($M_{\rm{WD}}^{i}$, $M_{2}^{i}$, $\rm{log_{10}}$\,$P^{i}$) of the close WD binaries obtained in our detailed 
binary evolution calculations. In our BPS calculations, the initial mass function of \citet{Mill79} is used. 
We assume a circular binary orbit and set up a constant initial mass ratio distribution (i.e. $n({q}')=$constant, 
see \citealt{Gold94, Bend08, Duch13}). The initial separation distribution of \citet{Han95} is adopted. 
We assume a constant star formation rate \citep{Han08}. The standard energy equations of \citet{Webb84} 
are used to calculate the output of the CE phase. The CE ejection is determined with two highly uncertain 
parameters, $\alpha_{\rm{CE}}$ and $\lambda$. Here, $\alpha_{\rm{CE}}$ is the CE ejection efficiency, i.e. the 
fraction of the released orbital energy used to eject the CE and $\lambda$ is a structure parameter that depends 
on the evolutionary stage of the donor star. In this work, the models with a metallicity of $Z=0.02$ and a parameter 
of $\alpha_{\rm{CE}}\lambda$=0.5 are adopted \citep{Han04}.

\begin{figure*}
  \begin{center}
    {\includegraphics[width=1.3\columnwidth, angle=270]{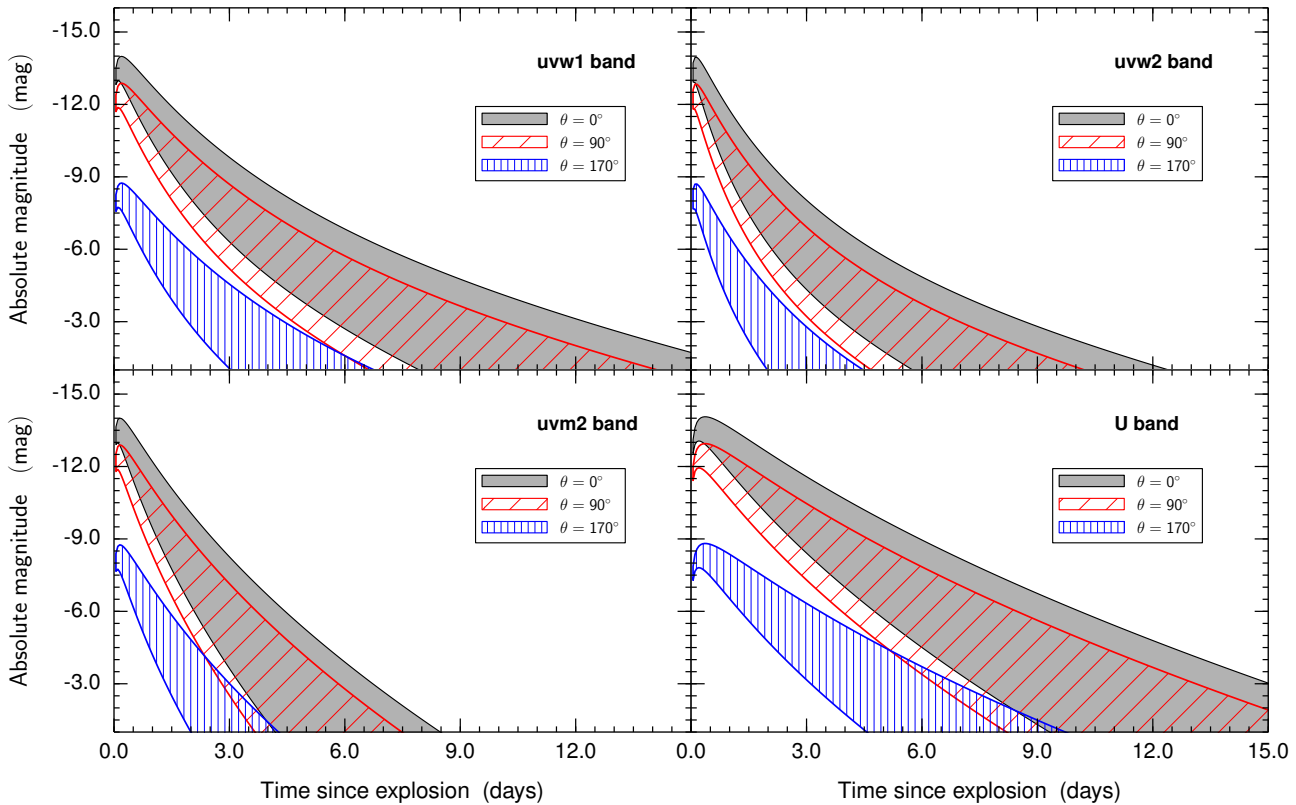}}
  \caption{As Fig.~\ref{Fig:1}, but for the He star donor Chandrasekhar-mass scenario.}
\label{Fig:3}
  \end{center}
\end{figure*}

\section{Predictions of the early UV emission}
\label{sec:3}

In the analytical model of \citet{Kase10}, SN ejecta collides with its companion star, the impacting layers 
are re-shocked and the kinetic energy partially is dissipated, causing UV flux emission from the shock-heated 
region. This emission dominates the early-time (few days after SN explosion) light curves of SNe Ia 
within certain viewing angles \citep{Kase10}. At later epochs, this shock flux becomes weak and the light 
curves of SNe Ia are dominated by the flux of the SN itself. \citet{Kase10} analytically estimated the early 
isotropic bolometric luminosity $L_{\rm{c,iso}}$ by the 
collision as (Equation~22 of \citealt{Kase10})
 \begin{equation}
%\small
\begin{split}
    \label{eq:1}
      L_{c,iso} = 10^{43}\left(\frac{a}{10^{13}\,\rm{cm}}\right)\left(\frac{M_{\rm{ej}}}{1.4\,M_{\sun}}\right)^{1/4}\left(\frac{v_{\rm{ej}}}{10^{\,9}\,\rm{cm\,s^{-1}}}\right)^{7/4}
      \\\times\left(\frac{\kappa_{e}}{0.2\,\rm{cm^{2}\,g^{-1}}}\right)^{-3/4}\left(\frac{t}{\rm{d}}\right)^{-1/2}\ \ \ \ \ \ \  \rm{erg\,s^{-1}},
\end{split}
   \end{equation}
where $a$ is the binary separation at the moment of SN explosion, $M_{\rm{ej}}$ is the SN ejecta 
mass, $v_{\rm{ej}}$ is the velocity of the SN ejecta colliding with the binary star, $\kappa_{e}$ is the 
electron scattering opacity in the SN ejecta which is assumed to be a constant value in this work, and $t$ is the time since SN explosion. 
This luminosity is strongly dependent on the binary separation at the moment of SN Ia explosion.
In addition, the effective temperature of the prompt shock emission is estimated as follows (Equation~25 of \citealt{Kase10}).
 \begin{equation}
%\small
\begin{split}
    \label{eq:2}
      T_{\rm{eff}} = 2.5\times10^{4}\left(\frac{a}{10^{13}\,\rm{cm}}\right)^{1/4}\left(\frac{\kappa_{e}}{0.2\,\rm{cm^{2}\,g^{-1}}}\right)^{-35/36}
      \\\times\left(\frac{t}{\rm{d}}\right)^{-37/72}\ \ \ \ \ \ \rm{K},
\end{split}
   \end{equation}

The strong UV emission at early times from the SN ejecta interacting with the companion star not only strongly 
depends on the progenitor system but also on the viewing angle $\theta$ \citep{Kase10}. For a given binary progenitor 
system, the maximum excess emission should be observed when the viewing angle is $0^{\circ}$, i.e., the companion 
star lies directly along the line of sight between the observer and the SN  explosion. On the contrary, the excess emission 
detected for viewing angle $\theta=180^{\circ}$ should be negligible, corresponding to a geometry in which the SN Ia lies
directly in the line of sight between the observer and the companion star. \citet{Brow12} estimated the dependence of the early 
emission from SN-companion interaction on the viewing angle. They found that the dependence of fractional flux values ($f$) on 
the viewing angle ($\theta$) can be fitted by a function (Equation~3 of \citealt{Brow12}) of the form:
 \begin{equation}
\begin{split}
    \label{eq:3}
     f = (0.5\,\rm{cos}\,\theta+0.5)\times(0.14\,\theta^{\,2}-0.4\,\theta+1),
\end{split}
   \end{equation}
where the viewing angle $\theta$ is in unit of radians. In this work, we directly adopt this 
function to calculate the observed flux at different viewing angles for a given epoch.

To facilitate direct comparison with the early-time observations of SNe Ia, the early isotropic bolometric luminosity 
of shocked gas is converted to broad band magnitudes by using the same method as \citet{Brow12}. Adopting the progenitor 
properties at the moment of SN Ia explosion obtained from our BPS calculations in Section~\ref{sec:2}, we create a 
set of blackbody models under a given viewing angle with appropriate temperature and luminosity based on  
Equations~\ref{eq:1}, \ref{eq:2} and \ref{eq:3}. In Fig.~\ref{Fig:sep}, we present the distributions of binary separation
at the moment of SN explosion in our BPS calculations for three different SD progenitor scenarios. It is shown that most progenitor systems
in the He star donor scenario have a relatively small separation (less than $10^{11}\,\rm{cm}$) at the moment of SN explosion, and the binary systems 
in the H-rich donor scenario cover a separation range of about $10^{11}-4.0\times10^{12}\,\rm{cm}$. With the created blackbody models, we then calculate 
the synthetic magnitudes at different bands by using transmission curves of chosen filters. Specifically for this work, 
the $\textit{Swift}$/UVOT filter system and their corresponding transmission curves are selected. The 
absolute magnitudes are calculated in the AB magnitude system.

\begin{figure*}
  \begin{center}
    {\includegraphics[width=0.33\textwidth, angle=270]{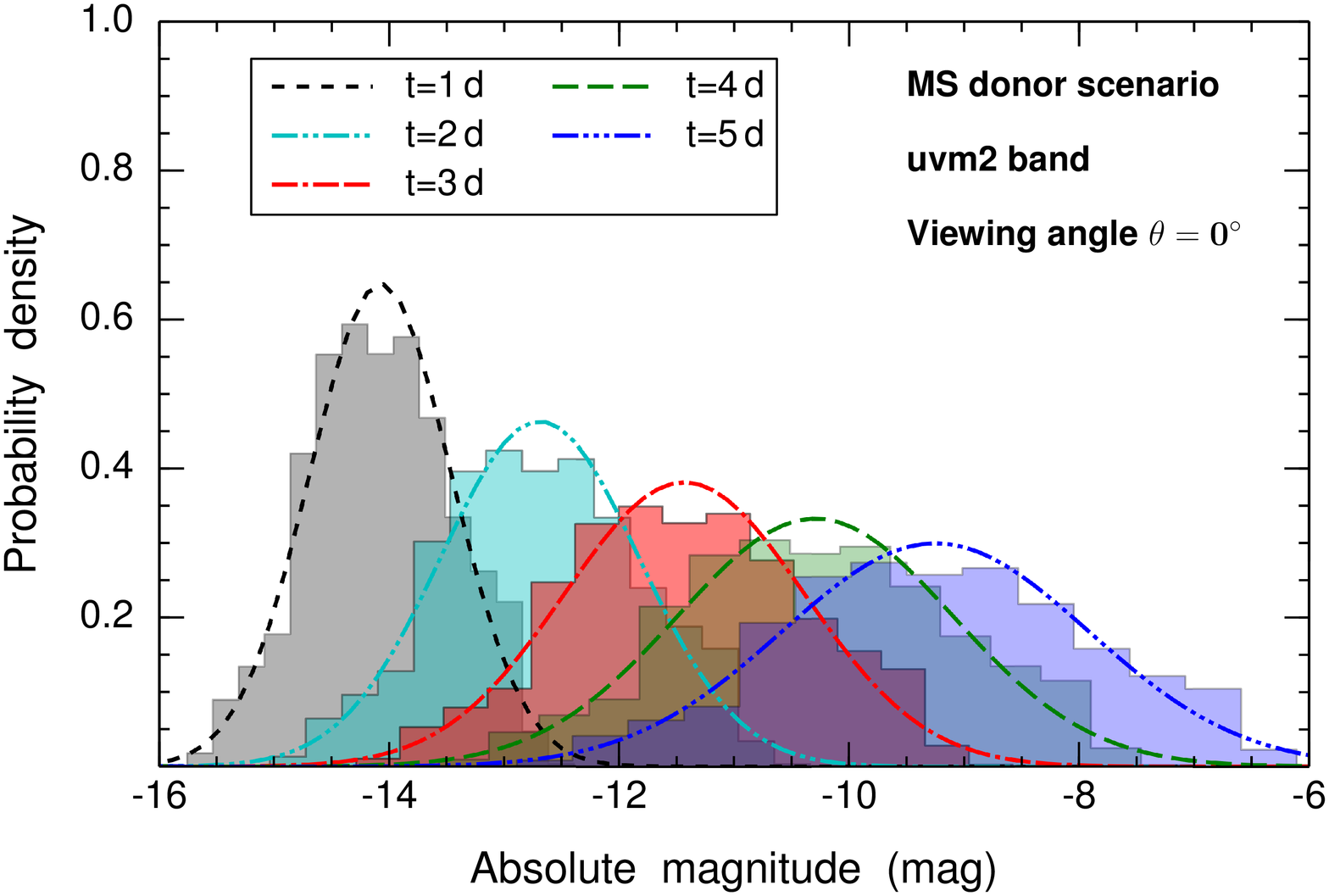}}
    \vspace{0.1in}
    {\includegraphics[width=0.33\textwidth, angle=270]{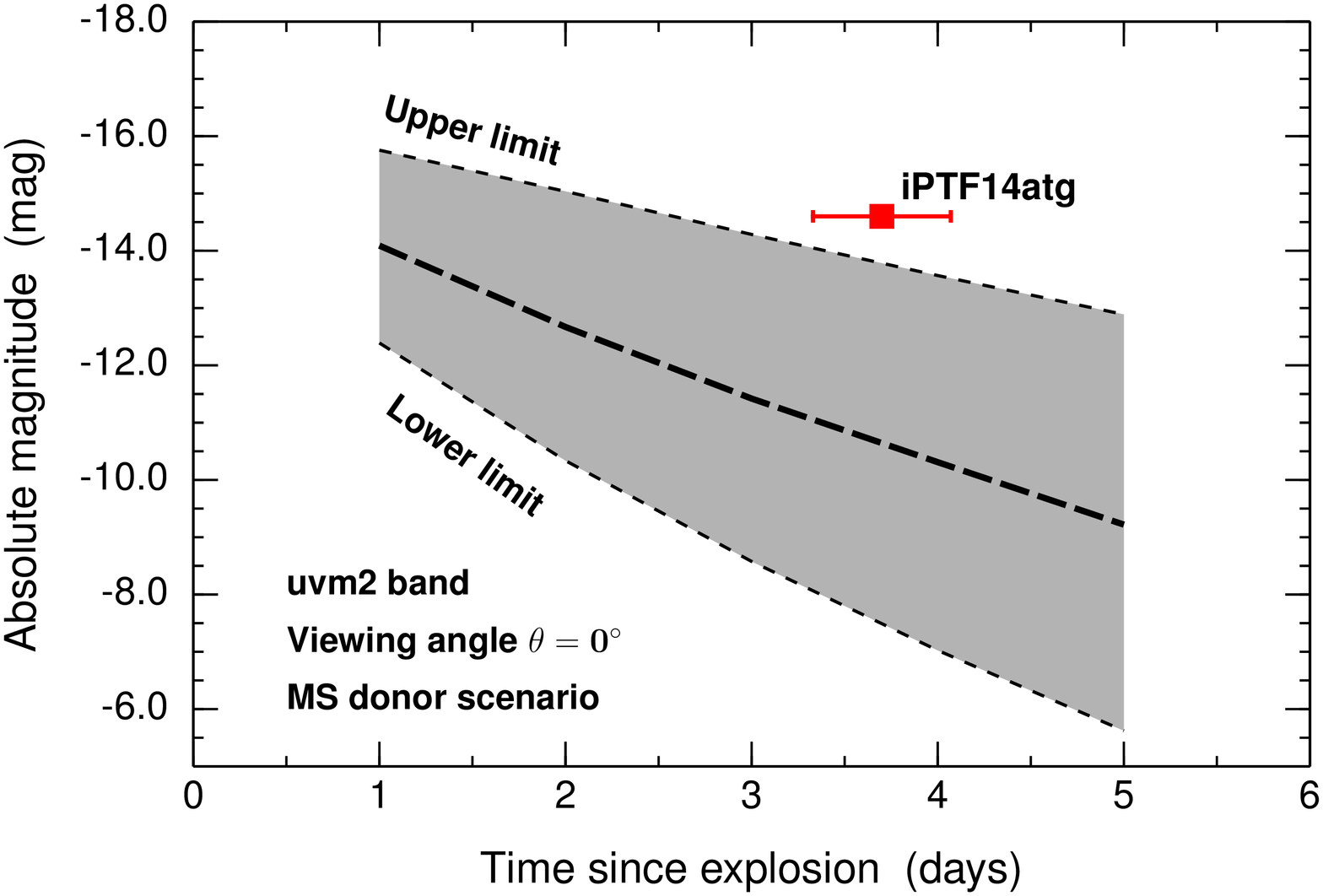}}
    \vspace{0.1in}
    {\includegraphics[width=0.33\textwidth, angle=270]{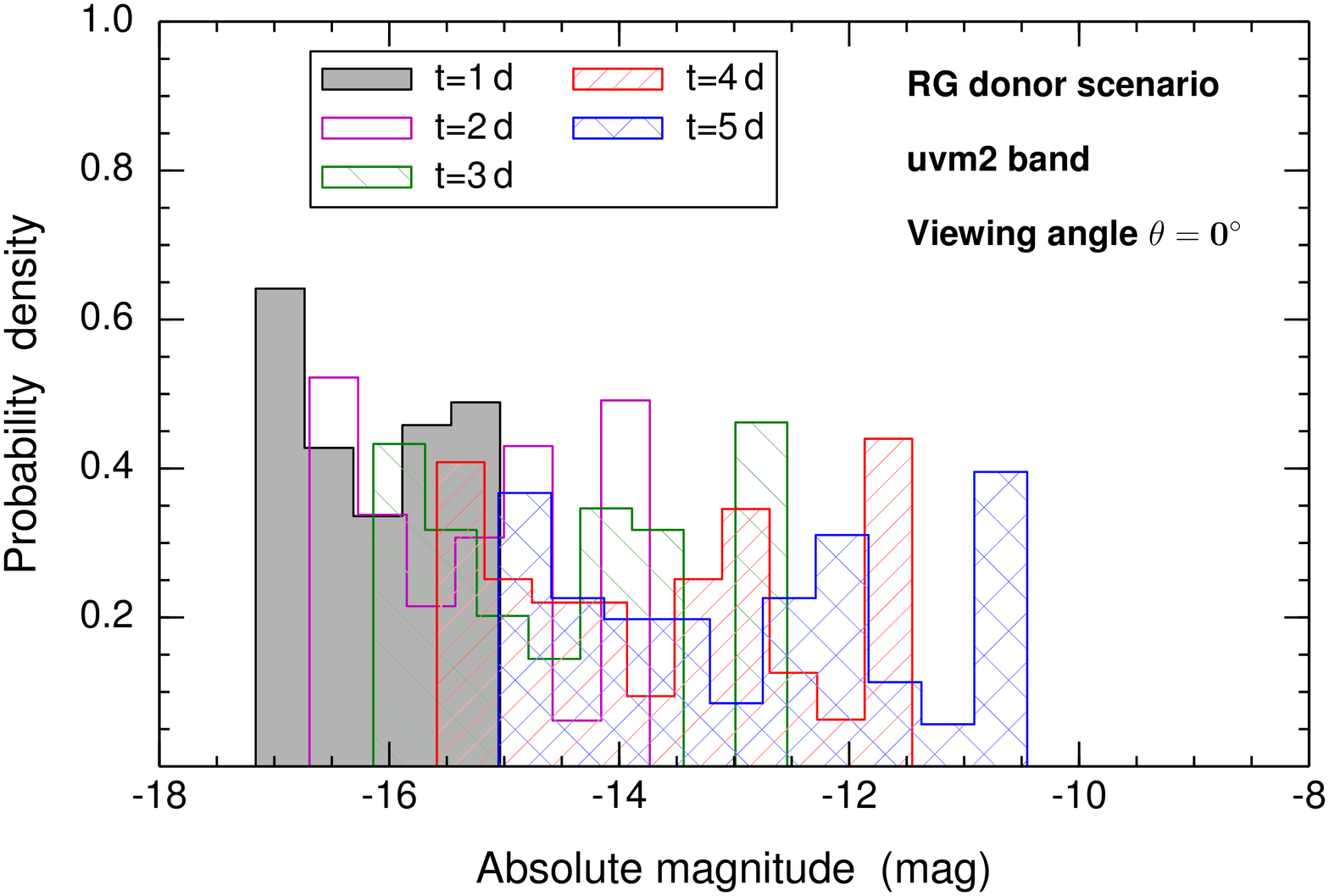}}
    \vspace{0.1in}
    {\includegraphics[width=0.33\textwidth, angle=270]{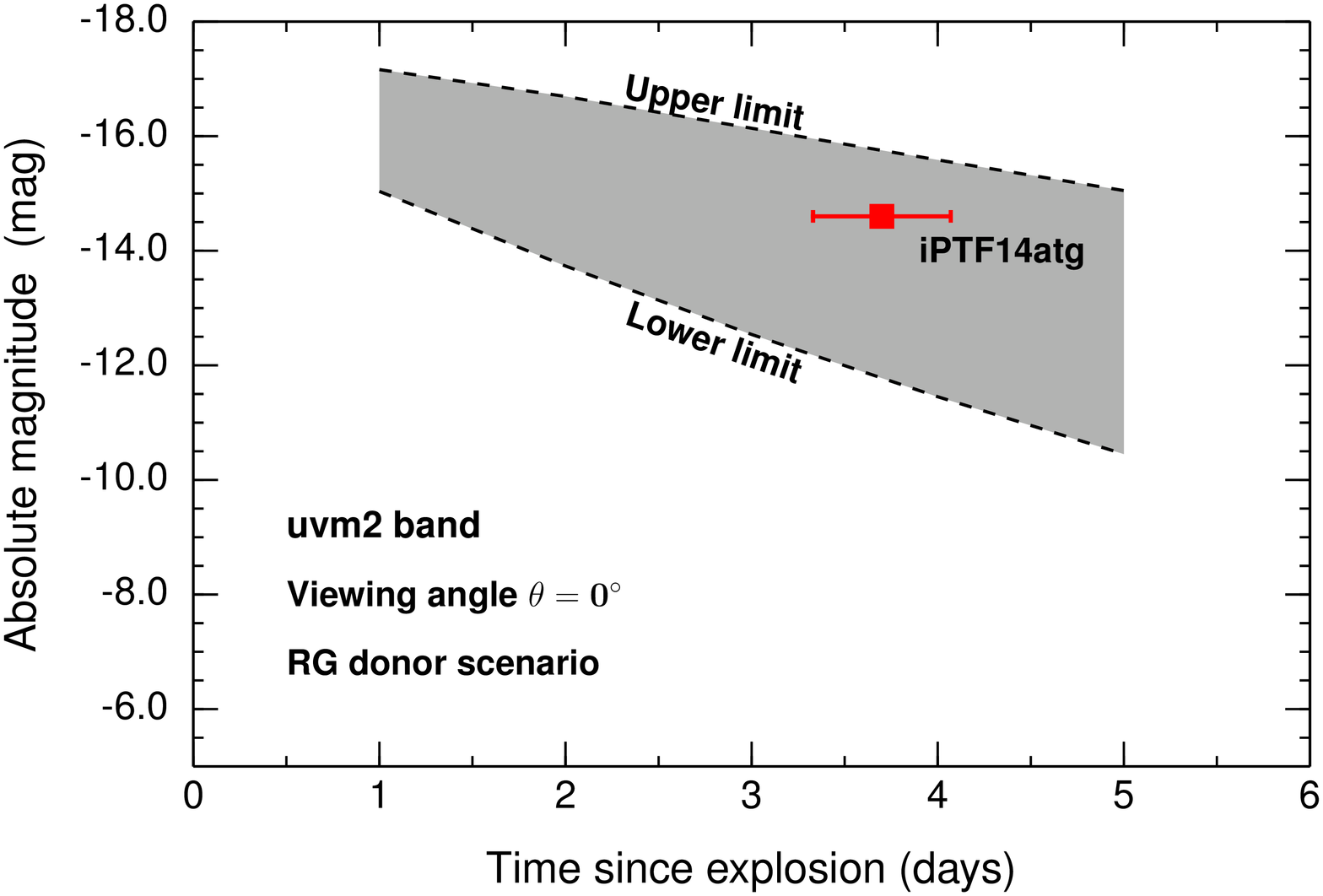}}
%    \vspace{0.1in}
    {\includegraphics[width=0.33\textwidth, angle=270]{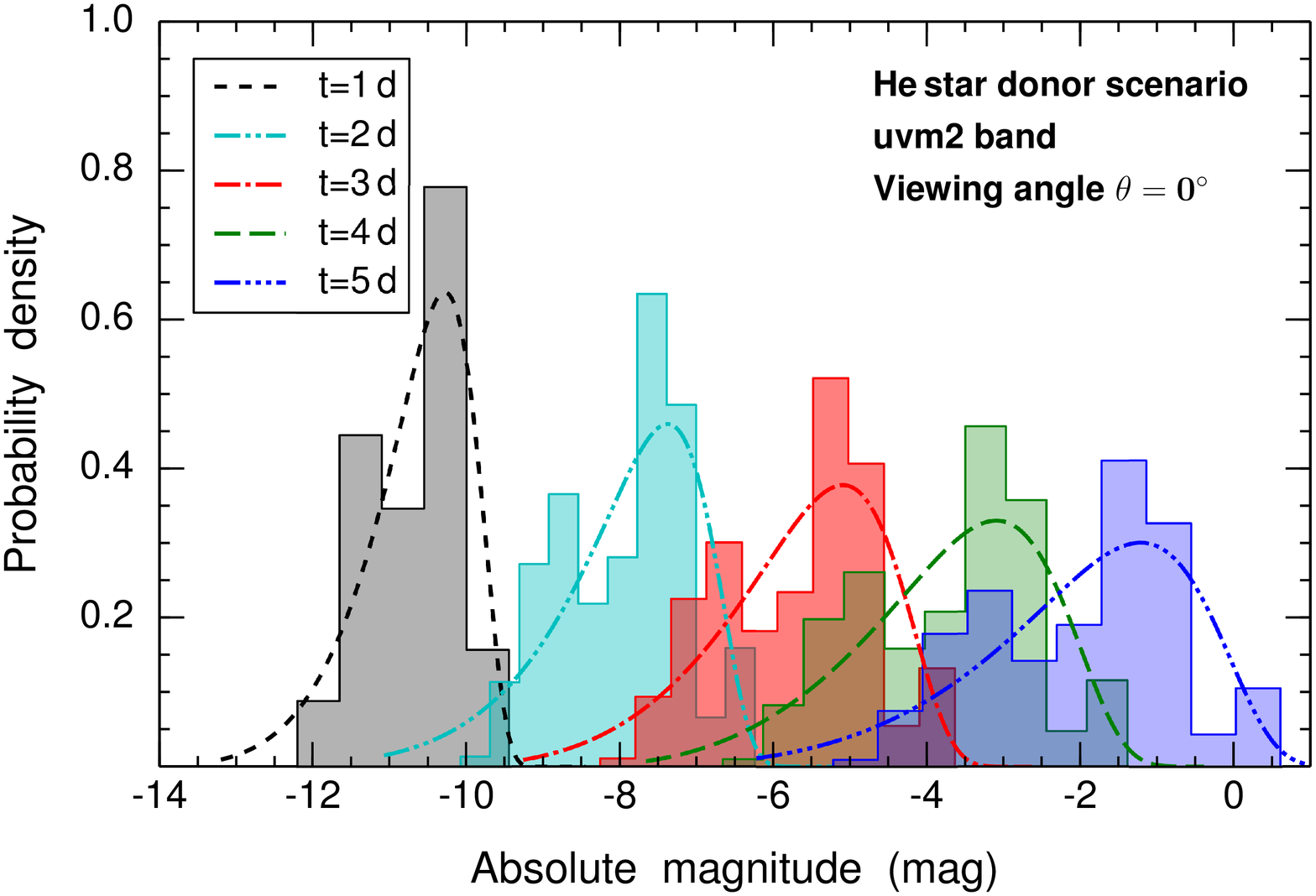}}
%    \vspace{0.1in}
    {\includegraphics[width=0.33\textwidth, angle=270]{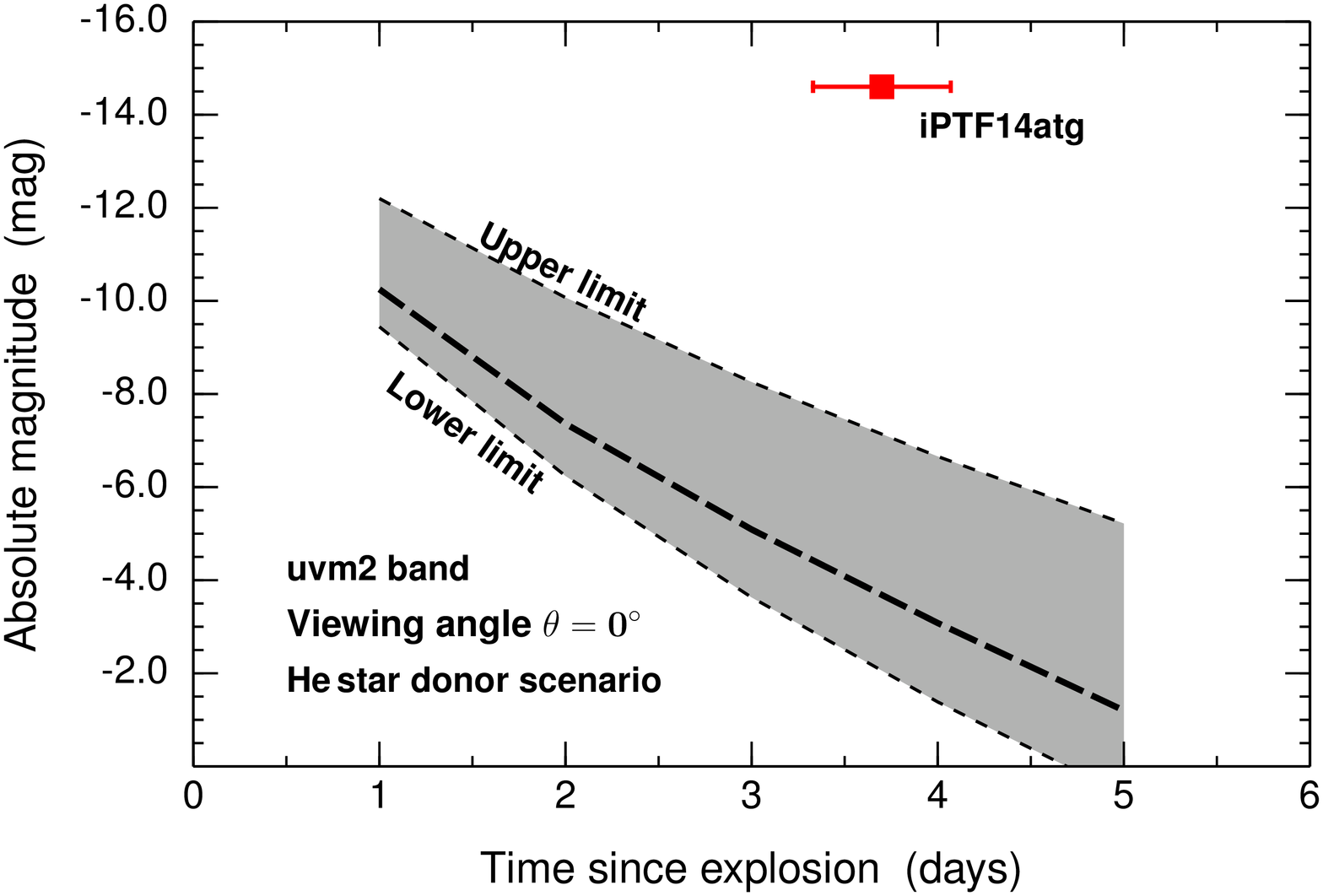}}
  \caption{Left: the probability distribution of the absolute magnitudes of shocked gas in $uvm2$ band at different epochs after the SN Ia explosion for
           the MS (top), RG (middle) and He star (bottom) donor scenario, respectively. The fitting curves of the distributions
           are also plotted except for the RG donor scenario. Right: comparison between the early observations of a peculiar SN Ia
           iPTF14atg and our predictions. The boundaries (short thin dashed curves) of gray region correspond to the upper-limit and lower-limit magnitude in $uvm2$ band.
           The long bold dashed lines show absolute magnitudes at the peaks of fitting curves of corresponding figures in left panel.
           The location of the early UV flash detected in iPTF14atg \citep{Cao15} is shown with the square in red simply assuming $10\%$ error of the time because 
           of the uncertainty of the explosion date. 
            }
\label{Fig:4}
  \end{center}
\end{figure*}

With different donor star models, the theoretical distributions of UV emission at early times from SN ejecta-companion interaction 
in the SD scenario are presented in Figs.~\ref{Fig:1},~\ref{Fig:2} and \ref{Fig:3}.
Here, we assume typical explosion parameters of an SN Ia, i.e., the amount of ejecta mass $M_{\rm{ej}}=1.4\,M_{\rm{\sun}}$ and
the SN explosion energy $E_{\rm{ej}}=1.0\times10^{51}\,\rm{erg}$. These typical values lead to a mean expansion velocity of SN
ejecta of $\approx10^{4}\,\rm{km\,s^{-1}}$. Here, only four different $\textit{Swift}$/UVOT filters are chosen to show as examples, 
i.e., the $uvw1$, $uvm2$, $uvm2$ and $U$ bands (the $B$ and $V$ band are not included), because the excess emission at early times should be brightest in the 
UV and become subordinate at longer optical wavelengths \citep{Kase10, Brow12}. For a given viewing angle, we show the upper-limit and lower-limit
absolute magnitude in each band as a function of time since the SN explosion. It is shown that the theoretical UV luminosities of shocked gas  
in the MS donor scenario cover a relatively wider magnitude range because binary systems in this scenario have a wide range of  progenitor 
properties at the moment of SN explosion compared to those in other two donor scenarios (Fig.~\ref{Fig:sep}). In addition, the early UV 
luminosities in the He star donor scenario are much weaker than those of the H-rich scenario by more than three magnitudes because 
He star donor progenitor systems have closer separations (Fig.~\ref{Fig:sep}) when the WDs grow to the Chandrasekhar-mass
limit (Fig.~\ref{Fig:3}). We note that the explosion parameters of typical SNe Ia used here may not match the case 
of iPTF14atg because it is identified as a subluminous SN~2002es-like event (see Section~\ref{sec:iPTF14atg}), which however,  
strongly depends on the exact explosion mechanism of a Chandrasekhar-mass CO WD.

In Figs.~\ref{Fig:1}--\ref{Fig:3}, we present possible magnitude ranges that can be covered by the early UV luminosities 
arising from the ejecta-companion interaction in different SD progenitor scenarios. To better compare our predictions 
with early-time UV observations of SNe Ia and put stronger constraints on the possible progenitors of SNe Ia, the probability 
distributions of our predicted UV luminoisties at a given epoch after the SN explosion are further shown in Fig.~\ref{Fig:4}. Here, we only consider
the case with a viewing angle $\theta=0^{\circ}$, i.e., the maximum flux case, and the results at $uvm2$ band are chosen to give
an example because $uvw1$ and $uvw2$ filters have an extended red tail. 
The distributions of different viewing angles and other bands 
can be roughly estimated according to the results in Fig.~\ref{Fig:1} to Fig.~\ref{Fig:4}. Also, only the results 
within the first five days after the SN explosion are presented in Fig.~\ref{Fig:4} because SN light curves at late times are dominated by the 
flux from the SN itself \citep{Kase10}. As it is shown in Fig.~\ref{Fig:4}, a peak is clearly seen in the probability 
distributions of the MS and He star donor scenario. However, the probability distributions of 
absolute magnitudes within a given day in the RG scenario is relatively flat (Fig.~\ref{Fig:4}). 
This may indicate that early UV luminosities of different progenitor systems in the RG donor scenario 
have a similar probability at a given epoch after SN explosion.

\section{Discussions}
\label{sec:4}

\subsection{Comparison with  the early UV flash of iPTF14atg}
\label{sec:iPTF14atg}

A strong UV pulse in the early-time light curves of a subluminous 
SN Ia iPTF14atg has been detected by \citet{Cao15}. They further suggested that this strong UV pulse
supports that some SNe Ia arise from the SD Chandrasekhar-mass scenario. Comparing their
observations to analytical models of \citet{Kase10}, they found that the theoretical UV emission from shocked 
ejecta of a binary system with a separation of about $60\,R_{\sun}$ can match
the observations assuming the explosion energy $E_{\rm{ej}}=10^{51}\,\rm{erg}$ and ejecta mass 
$M_{\rm{ej}}=1.4\,M_{\sun}$. Here, we also compare the predictions from our BPS calculations for
different SD scenarios to the early UV flash of iPTF14atg to put a more stringent on 
the possible progenitor of this event.

A detailed comparison with the results of the $uvm2$ band is given in the right column of Fig.~\ref{Fig:4}. 
It shows that the observed early UV pulse of iPTF14atg is stronger 
than the upper limit of the theoretical UV luminosity predicted from the MS donor and He 
star donor scenario at the same epoch (Fig.~\ref{Fig:4}). Considering that the actual viewing 
angle may be larger than $\theta=0^{\circ}$ and taking the uncertainties on the explosion 
parameters (see Section~\ref{sec:models}) into account, it is unlikely that the detected UV flash 
in early-time light curves of iPTF14atg arises from the interaction of SN Ia ejecta with a MS or 
He star companion. However, our theoretical early UV luminosities from the RG donor scenario 
can include the early UV flash of iPTF14atg (Fig.~\ref{Fig:4}) although the theoretical upper-limit luminosity 
in this scenario can still decrease to a value weaker than the observed UV flash in iPTF14atg as the viewing angle increases.  
Therefore, whether the predictions from the RG donor scenario can support the early UV observations of iPTF14atg 
depends on the actual viewing angle of the observations. Nevertheless, we still suggest that the 
RG donor binary system as progenitor of iPTF14atg is more possible if the early UV flash detected in an SN~2002es-like SN iPTF14atg
indeed arises from the ejecta-companion interaction in the SD scenario.

\begin{figure}
  \begin{center}
    {\includegraphics[width=0.77\columnwidth, angle=270]{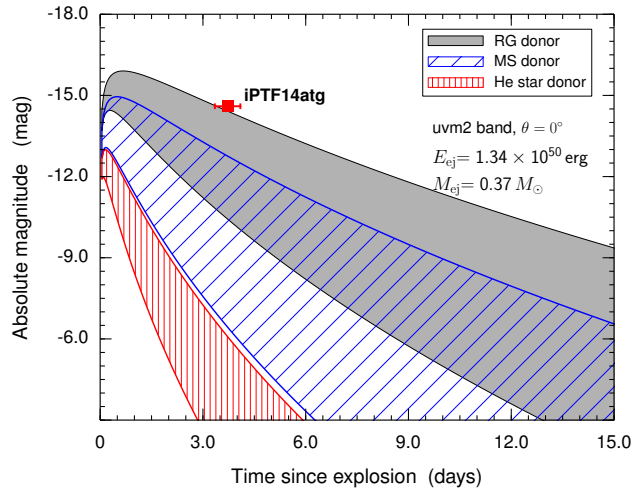}}
  \caption{Similar as Fig.~\ref{Fig:1}, but for theoretical predictions of 
            SN Iax case. Here, only a viewing angle $\theta=0^{\circ}$ and $uvm2$ band are considered.
           The ejecta mass ($M_{\rm{ej}}=0.37\,M_{\sun}$) and SN explosion 
           energy ($E_{\rm{ej}}=1.34\times10^{50}\,\rm{erg}$) are directly derived from a weak deflagration 
           explosion of the Chandrasekhar-mass C/O WD for SN Iax event \citep{Krom13}. The location of early UV flash of iPTF14atg
           is also shown \citep{Cao15}.}
\label{Fig:5}
  \end{center}
\end{figure}

\begin{figure*}
  \begin{center}
    {\includegraphics[width=0.33\textwidth, angle=270]{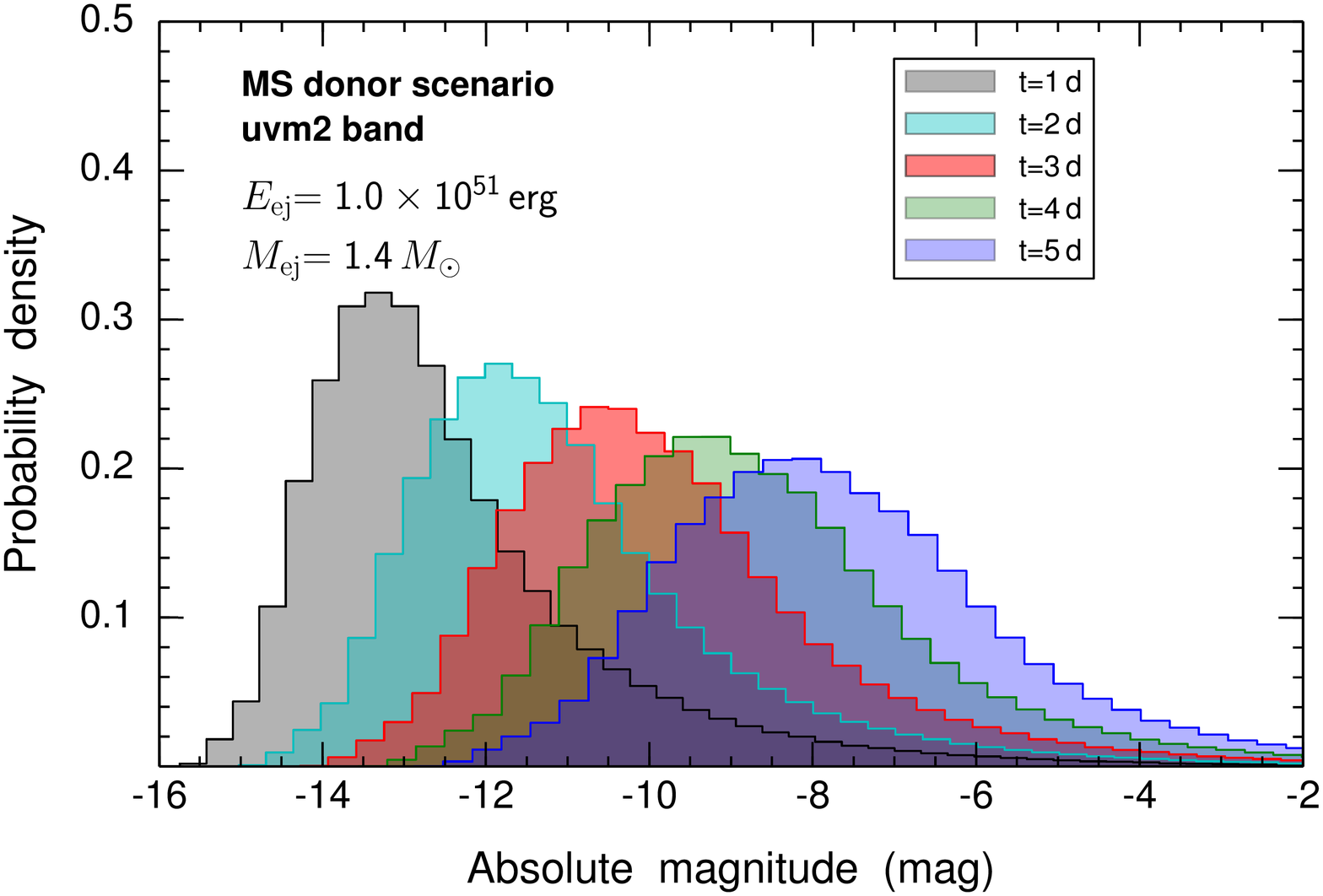}}
    \vspace{0.1in}
    {\includegraphics[width=0.33\textwidth, angle=270]{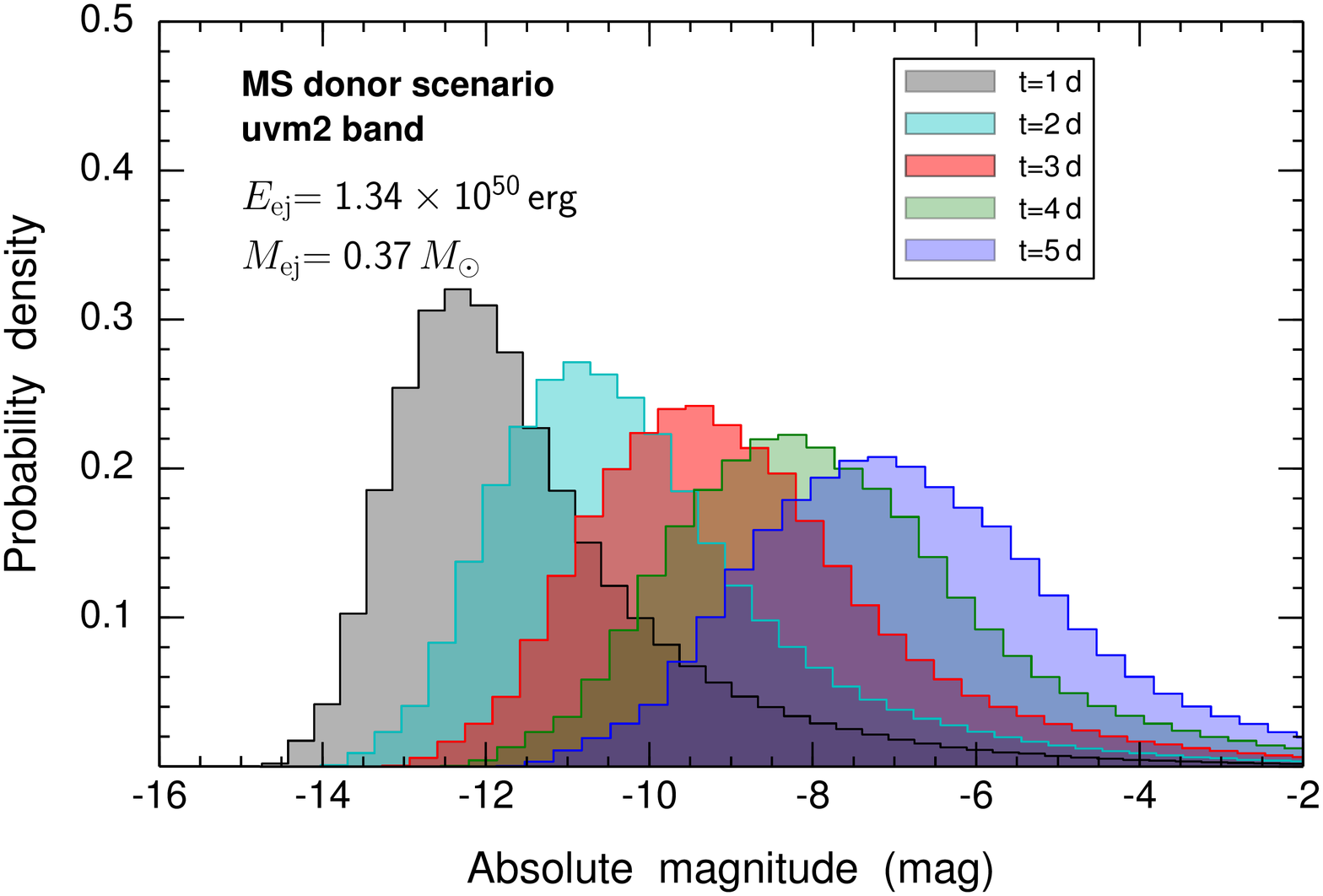}}
    \vspace{0.1in}
    {\includegraphics[width=0.33\textwidth, angle=270]{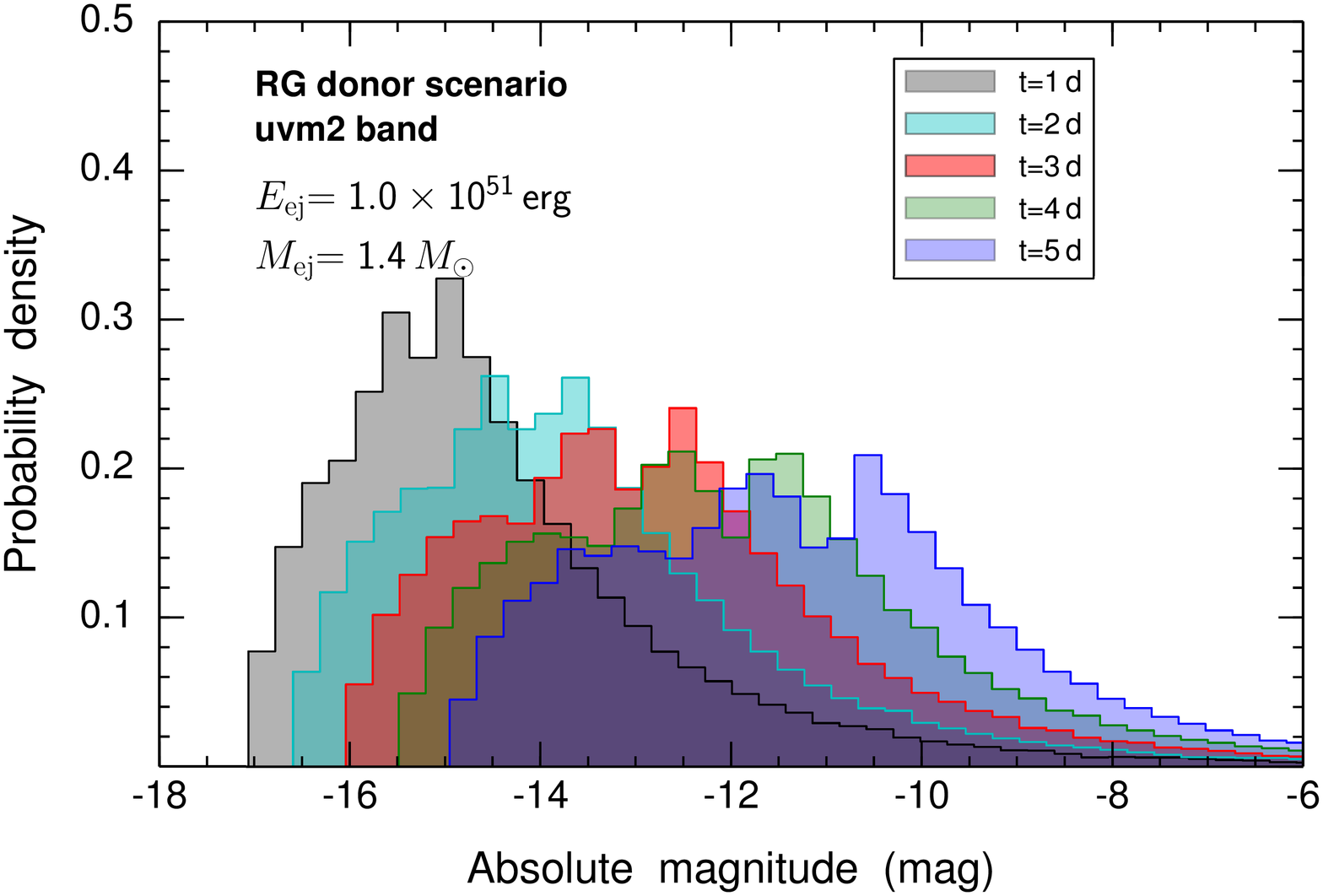}}
    \vspace{0.1in}
    {\includegraphics[width=0.33\textwidth, angle=270]{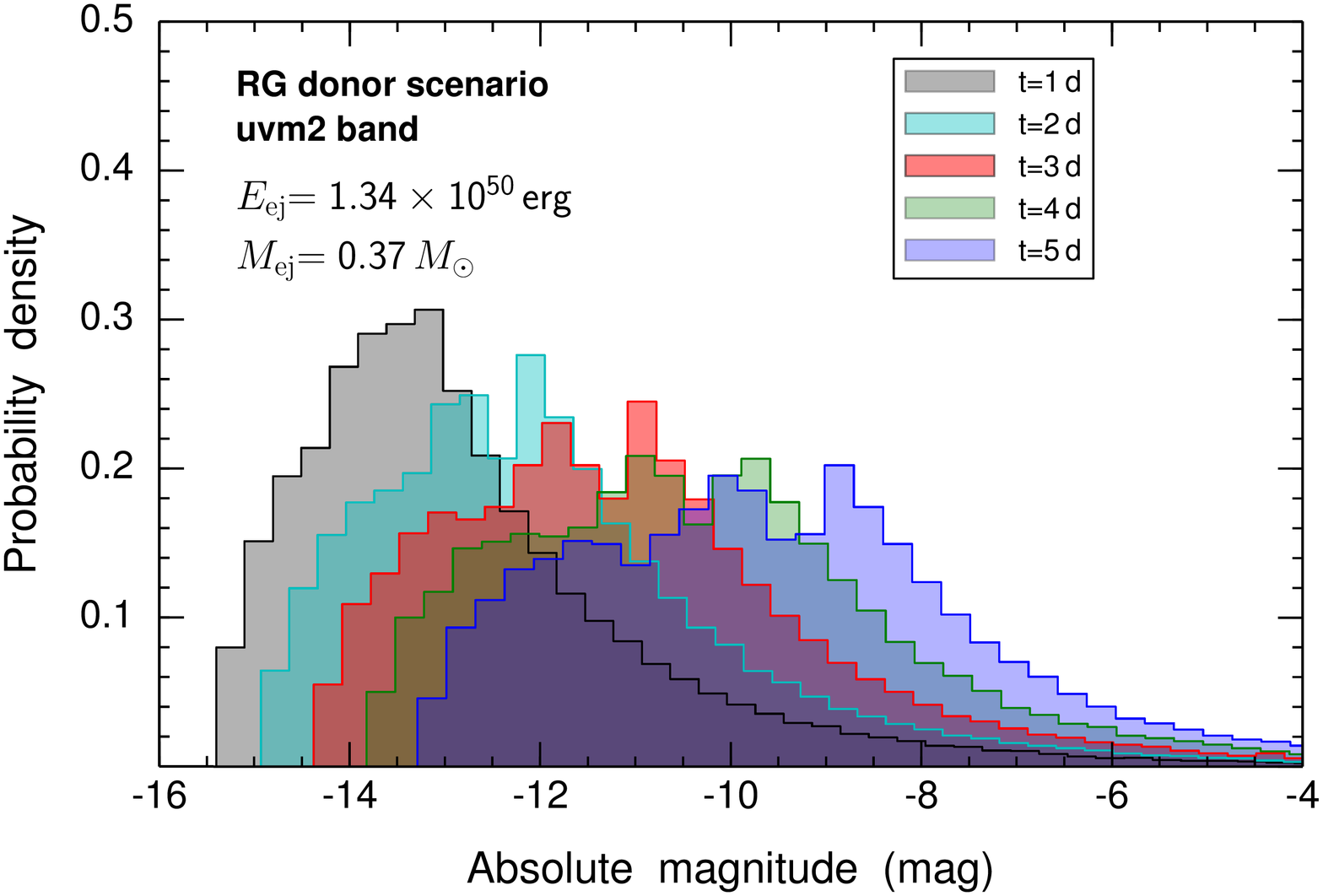}}
%    \vspace{0.1in}
    {\includegraphics[width=0.33\textwidth, angle=270]{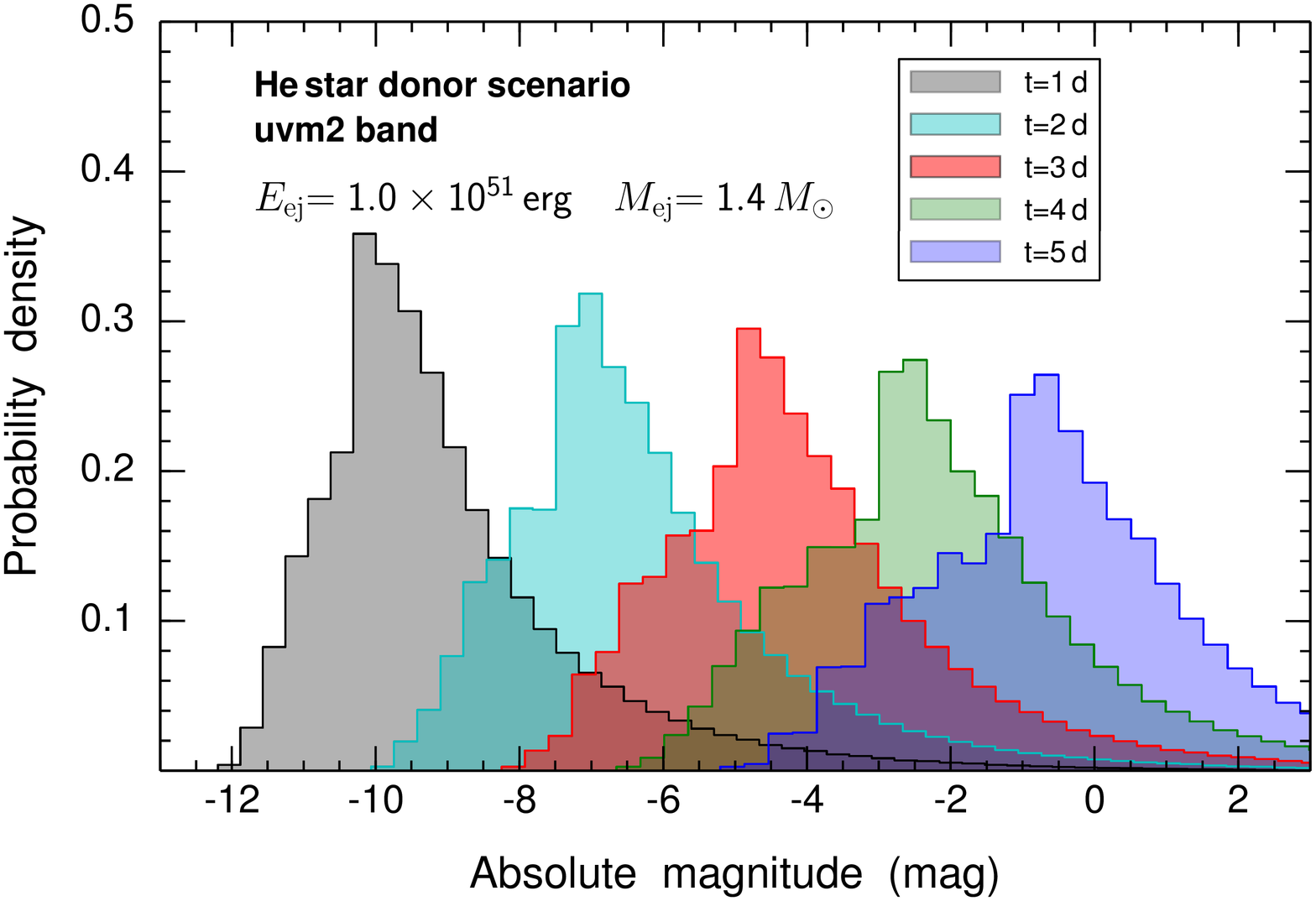}}
%    \vspace{0.1in}
    {\includegraphics[width=0.33\textwidth, angle=270]{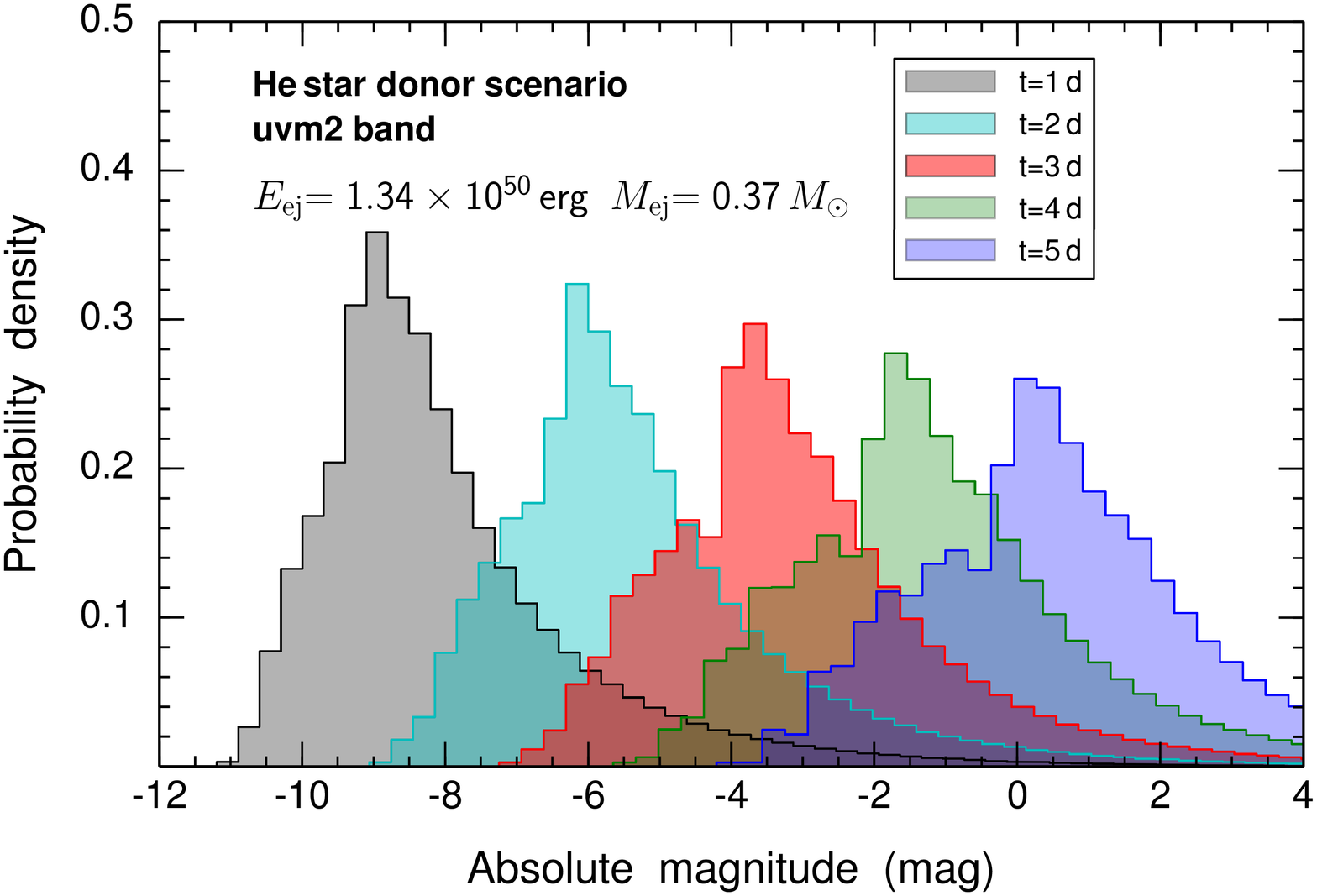}}
  \caption{Similar to Fig.~\ref{Fig:4}, but showing the results with a random distribution of viewing angle $\theta$. The distributions by adopting the explosion parameters
           for typical SNe Ia (left column) and SN Iax event (right column) are presented.  
            }
\label{Fig:6}
  \end{center}
\end{figure*}

\subsubsection{Different explosion parameters}
\label{sec:Iax}

The comparisons given above are based on an assumption of the typical SN Ia explosion parameters as those used in
\citet{Cao15}. However, it has been suggested that iPTF14atg is more likely to belong to the subluminous SN~2002es family by \citet{Cao15}. Meanwhile, it is
shown that the peak magnitude of iPTF14atg is similar to SN~2005hk (a typical SN Iax event). 
The hydrodynamical simulations of \citet{Krom13} have shown that the off-centre-ignited weak deflagration of Chandrasekhar-mass CO WDs is 
able to reproduce the characteristic observational features of SN~2005hk quite well. Adopting the 
weak deflagration explosion model ($E_{\rm{ej}}=1.34\times10^{50}\,\rm{erg}$ and $M_{\rm{ej}}=0.37\,M_{\sun}$, which corresponds to 
a mean expansion velocity $v_{\rm{ej}}\approx6\,000\,\rm{km\,s^{-1}}$) of \citet{Krom13}
as inputs of the analytical model of \citet{Kase10}, we predict theoretical early UV luminosities in the $uvm2$ band of
different SD progenitor scenarios (Fig.~\ref{Fig:5}). Compared to the results with typical explosion 
parameters in Figs.~\ref{Fig:1}--\ref{Fig:3}, it is shown that the early UV luminosities of shocked ejecta are weaker 
by about one magnitude when the explosion parameters of Iax events are used (Fig.~\ref{Fig:5}). If the SN iPTF14atg 
was produced from an explosion model that has similar explosion parameters 
with the weak deflagration explosion model for SNe Iax in \citet{Krom13}, the observed UV flash 
in iPTF14atg is a bit stronger than the upper-limit luminosity of our RG donor model at similar 
epoch (Fig.~\ref{Fig:5}). 

However, the actual explosion date was not well constrained in \citet{Cao15}. Therefore, the 
delay time between the SN explosion and the moment of ejecta-companion interaction happens 
is not well determined, which means that the location of early UV pulse of iPTF14atg in
Figs.~\ref{Fig:4} and ~\ref{Fig:5} can be shifted a bit towards the left (or the right) if the actual explosion date
is earlier (or later) than that used in \citet{Cao15}.

\subsubsection{SN birth rate}
\label{sec:rate}

In our BPS calculations, we obtain a SN Ia rate of about $3\times10^{-5}\,\rm{yr^{-1}}$ (i.e., three 
SN Ia events every $10^{5}$ years) in the RG donor channel for a constant star
formation rate ($5.0\;\rm{M_{\sun}\,yr^{-1}}$), which 
is about $1\%$ of the total SN Ia rate if we assume that the total SN Ia rate is $3.5\times10^{-3}\,\rm{yr^{-1}}$  based on 
the inferred Galactic SN Ia rate of $3-4\times10^{-3}\,\rm{yr^{-1}}$ \citep{Capp97}. Observationally, \citet{Gane12} estimate 
that roughly $2.5\%$ of SNe Ia may belong to SN~2002es-like events within a fixed volume. The birth rate from the RG donor
channel in our BPS calculation is roughly comparable with observed rate of SN~2002es-like event (but see \citealt{Hach99}). 
In addition, the delay times between star formation and SN explosion in the RG donor scenario are quite 
long ($\gtrsim3\,\rm{Gyr}$), which seems to support the fact that 2002es-like SNe Ia are likely from 
an old stellar population \citep{Gane12} and the host galaxy of iPTF14atg is classified as an old galaxy.    

However, the contribution of the RG donor scenario to the total SN Ia rate is quite uncertain \citep{Hach99, Han04, Chen11}. If 
mass transfer in symbiotic binaries (RG donor model) occurs through the stellar wind rather than the RLOF assumed in this 
work \citep{Bran95}, the SN contribution from this scenario could remarkably increase and binary systems could have much larger orbital separations
than those in this work at the moment of SN Ia explosion \citep{Chen11}. Subsequently, early UV emissions from shocked gas may be much
stronger than those obtained from our RG donor models.

\subsection{The effect of viewing angle}

In Fig.~\ref{Fig:4}, we present the expected magnitudes in the $uvm2$ band from a fixed viewing 
angle ($\theta=0^{\circ}$). However, we do not necessarily observe the interaction from $\theta=0^{\circ}$.
To investigate the effect of the viewing angle on the observational magnitude,
we randomly assign the viewing angle and obtain the probability distribution for the observational 
magnitude in the $uvm2$ band. Fig.~\ref{Fig:6} shows the probability distribution for both normal SNe 
Ia (i.e. $E_{\rm{ej}}=1.0\times10^{51}\,\rm{erg}$ and $M_{\rm{ej}}=1.4\,M_{\sun}$) and SN Iax 
events (i.e. $E_{\rm{ej}}=1.34\times10^{50}\,\rm{erg}$ and $M_{\rm{ej}}=0.37\,M_{\sun}$). Comparing to the distributions in Fig.~\ref{Fig:4},
the peak of the distribution becomes about one magnitude fainter after taking the effect of the viewing angle into account.
However, the maximum magnitude achieved by the SN ejecta-companion interaction does not change and our conclusion regarding the SN iPTF14atg 
remains unaltered.

\subsection{The Spin-up/Spin-down model}

In the SD scenario, a WD accretes and retains companion matter that
carries angular momentum. As a consequence the WD spins with a short
period which leads to an increase of the critical explosion mass. If the critical
mass is higher than the actual mass of the WD, the SN explosion can only
occur after the WD spins down to decrease its critical explosion mass to fall 
below the current mass of the WD with a specific spin-down
timescale \citep{Di11, Just11, Hach12}. In such case, the MS or RG donor star 
might shrink rapidly before the SN Ia explosion occurs because most its H-rich 
envelope has been exhausted during a long spin-down phase of the rapidly 
rotating WD. As a result, the early UV signature from SN Ia ejecta interacting 
with its companion star should be remarkably reduced compared to the results 
we presented in this work because the donor star
is much smaller than its Roche lobe at the moment of SN Ia explosion. In addition, as previously mentioned, this 
``spin-up/spin-down'' model seems to be able to explain the lack of H in late spectra
of SNe Ia and possibly the absence of  a surviving companion star in the SN remnants \citep{Di11, Just11, Hach12}.
However, the exact spin-down timescale of the WD in this model is quite unknown.

\subsection{The effect of different explosion models}
\label{sec:models}

In the SD scenario, depending on the exact ignition condition, a Chandrasekhar-mass CO WD can undergo 
a deflagration, a detonation, or a delayed detonation explosion \citep{Arne69, Woos86, Khok89, Roep05, Roep07}, which 
could finally produce SN Iax events \citep{Jord12b, Krom13, Krom15, Fink14}, normal SNe Ia \citep{Seit13}, or overluminous 
SNe Ia \citep{Plew04, Jord12a}. The distributions of early UV emission of shocked gas in normal SNe Ia
and Iax events have been presented 
in Sections~\ref{sec:3} and~\ref{sec:iPTF14atg}. 

By analytically deriving the existence of a characteristic length scale which establishes 
a transition from central ignitions to buoyancy-driven ignitions, \citet{Fish15} suggested 
that the Chandrasekhar-mass WDs in the SD scenario are generally expected to explode as 
overluminous SNe Ia rather than normal SNe Ia or Iax events. 
If the suggestion of \citet{Fish15} is true, early UV signatures due to the ejecta-companion interaction may 
be easier to be detected in overluminous SNe Ia because overluminous SNe Ia may have a higher explosion 
energy than that of normal SNe Ia. A higher explosion energy is expected to cause stronger early UV emissions 
of shocked gas for a given binary progenitor.

\subsection{Model uncertainties}

The prescription of the optically thick wind model of \citet{Hach96} and 
He-retention efficiencies from \citet{Kato04} are used to describe the mass accumulation 
efficiency of accreting WDs in our detailed binary evolution in this work. However, different 
mass-retention efficiencies may lead to the results that are different from our 
BPS calculations (e.g., see \citealt{Bour13, Pier14, Ruit14, Toon14}).
However, the exact mass-retention efficiency is still not well-constrained.

It should be kept in mind that the initial conditions of BPS calculations such as 
the CE evolution, current star-formation rate and initial mass function, may be 
sensitive to the assumed parameters in specific BPS codes, which lead to some uncertainties on 
the BPS results \citep{Toon14}. However, current constraints on these parameters (e.g., the CE efficiency, see \citealt{Zoro10, De11, Ivan13}) 
are still weak. For a detailed discussion for the effect of different theoretical 
uncertainties, see \citet{Clae14}. 

Finally, we point out that the analytical model of \citet{Kase10} is simple, 
which may lead to the predicted early UV luminosity being overestimated, 
or underestimated. For instance, a constant opacity from electron 
scattering is assumed in the analytical model of \citet{Kase10}, the luminosity can be reduced if the 
realistic opacity in the interacting material is decreased \citep{Brow12}. To put stronger constraints on the possible progenitor of 
SNe Ia, the radiation hydrodynamical simulations of the interaction
between SN Ia ejecta and its companion star are encouraged to address in the 
future study.

\section{Conclusion and Summary}

\label{sec:5}

In the SD Chandrasekhar-mass scenario of SNe Ia, strong UV emissions arising from the SN ejecta 
interacting with its companion star is suggested to be dominant in the early-time light curves
of SNe Ia \citep{Kase10}. These prompt UV emission characteristics are expected to constrain 
possible progenitor systems of SNe Ia. In this work, we perform BPS calculations for different 
progenitors of SNe Ia within the SD Chandrasekhar-mass scenario to obtain progenitor properties 
at the moment of SN Ia explosion. We then predict the distributions of early-time UV luminosities
of SN ejecta-companion interaction of different progenitors by combing our BPS results with 
the analytical model of \citet{Kase10}. These theoretical predictions for early UV 
emissions of shocked ejecta will be helpful to compare with future early-time observations of SNe 
Ia to put constraints on their possible progenitors.  

In addition, assuming the observed strong UV pulse of iPTF14atg indeed arises from the SN 
ejecta-companion interaction, we have compared our theoretical UV luminosities to the observed UV pulse at early times
of a peculiar subluminous SN Ia iPTF14atg \citep{Cao15}. With our BPS models, we find that 
predicted UV luminoisties in the MS and He star donor scenario struggle to reach an upper-limit absolute 
magnitude as strong as early UV flash observed in iPTF14atg, which suggests that the MS and He star donor 
binary systems are unlikely to be the progenitor. In particular, the He star donor binary system as 
progenitor of SN~iPTF14atg can be ruled out by our population synthesis calculations. However, theoretical UV luminosities 
at early times in our RG donor scenario provide a match with the early UV observations of iPTF14atg under some
appropriate conditions such as a favorable viewing angle. Therefore, we suggest that a red-giant donor binary system 
as the progenitor of an SN~2002es-like peculiar SN Ia iPTF14atg is more possible based on our populations synthesis model.

\section*{Acknowledgments}
     We thank the anonymous referee for his/her valuable comments and suggestions that helped to improve 
      the paper.  
      Z.W.L thanks Bo Wang for his helpful discussions about BPS calculations.
      We would also like to thank Carlo Abate and Norberto Castro for their helpful discussions.
      This work is supported by the Alexander von Humboldt Foundation. 
      R.J.S. is the recipient of a Sofja Kovalevskaja Award from the Alexander von Humboldt Foundation.
      T.J.M. is supported by Japan Society for the Promotion of Science
      Postdoctoral Fellowships for Research Abroad (26\textperiodcentered 51).

\bibliographystyle{mn2e}

\bibliography{ref}

\end{document}